\documentclass[twocolumn,aps,prl]{revtex4-2}

\usepackage{xcolor}
\usepackage{overpic}
\usepackage{amssymb}
\usepackage{graphicx}
\usepackage{times}
\usepackage{rotating}
\usepackage{float}
\usepackage{amsmath}

\begin{document} 


\title{Noisy intermediate-scale quantum computing algorithm for solving an $n$-vertex MaxCut problem with log($n$) qubits.}

\author{Marko J. Ran\v{c}i\'{c}}

\email{marko.rancic@totalenergies.com}

\address{TotalEnergies, Nano-INNOV – Bât. 861, 8, Boulevard Thomas Gobert – 91120 Palaiseau – France}

\date{\today}

\begin{abstract}
Quantum computers are devices which allow more efficient solutions of problems as compared to their classical counterparts. As the timeline to developing a quantum-error corrected computer is unclear, the quantum computing community has dedicated much attention to developing algorithms for currently available Noisy Intermediate-Scale Quantum computers (NISQ). Thus far, within NISQ, optimization problems are one of the most commonly studied and are quite often tackled with the Quantum Approximate Optimization Algorithm (QAOA). This algorithm is best known for computing graph partitions with a maximal separation of edges (\textit{MaxCut}), but can easily calculate other problems related to graphs. Here, I present a novel quantum optimization algorithm which uses exponentially less qubits as compared to the QAOA while requiring a significantly reduced number of quantum operations to solve the \textit{MaxCut} problem. Such an improved performance allowed me to partition graphs with 32 nodes on publicly available 5 qubit gate-based quantum computers without any preprocessing such as division of the graph into smaller subgraphs. These results represent a $40\%$ increase in graph size as compared to state-of-art experiments on gate-based quantum computers such as Google Sycamore. The obtained lower bound is $54.9\%$ on the solution for actual hardware benchmarks and $77.6\%$ on ideal simulators of quantum computers. Furthermore, large-scale optimization problems represented by graphs of a 128 nodes are tackled with simulators of quantum computers, again without any pre-division into smaller subproblems and a lower solution bound of $67.9 \%$ is achieved. The work presented here paves way to using powerful genetic optimizer in synergy with quantum computers.
	\end{abstract}
\maketitle 
\section{Introduction}
A universal quantum computer has been the holy grail of quantum technology \cite{feynman1982simulating}. Such a device would allow more efficient searching through databases \cite{grover1996fast}, prime number factorization \cite{shor1994algorithms}, and more efficient solutions of systems of linear equations \cite{PhysRevLett.103.150502}, just to name a few. However, universal quantum computers require millions of qubits with quantum error correction implemented and with an implementation timeline which is difficult to predict \cite{IBMQ_roadmap}. On the other hand, devices with up to 127 noisy qubits are readily available. This steered the scientific community towards exploring potential computational advantages which such devices could bring. In the rapidly expanding field of Noisy Intermediate-Scale Quantum Computing (NISQ) \cite{mcclean2016theory,preskill2018quantum} two algorithms stand out in prospect: the Variational Quantum Eigensolver (VQE)\cite{peruzzo2014variational, tilly2022variational} and the aforementioned Quantum Approximate Optimization Algorithm (QAOA) \cite{farhi2014quantum,cerezo2021variational,PhysRevX.10.021067}. The VQE is mainly applied to problems in chemistry and material science while the QAOA is best known for computing graph partitions with a maximal separation of edges (\textit{MaxCut}), but can easily calculate other properties of graphs, such as MaxIndependent Set and the partition problem, just to name a few. Given a graph $G=(V,E)$ comprising of $|V|$ vertices and $|E|$ nodes the QAOA requires $n=|V|$ qubits and $p\left(|E|+ |V|\right)$ quantum operations to implement the ansatz for the \textit{MaxCut} problem \cite{PhysRevX.10.021067,cerezo2021variational}. Here, $p$ is the phenomenological depth parameter.  

In this manuscript  a novel variational \textit{MaxCut} algorithm requiring $n=\lceil\log_2|V|\rceil$ qubits is introduced, where $\lceil$ $\rceil$ stands for the ceiling function. For example if $x=2.1$, $\lceil x\rceil=3$, if $x=2.7$, $\lceil x\rceil=3$. In similarity with all other NISQ algorithms the algorithm presented here iteratively improves a trial solution (ansatz) in a hybrid quantum-classical optimization loop. The ansatz is implemented with at most $2^n-2n+5$ single qubit gates and at most $2^n-2$ two qubit CNOT gates (in total up to $2^{n+1}-2n+3$ gates). Exploiting the fact that large graphs can be treated with the algorithm at a low resource overhead I demonstrate the calculation of a \textit{MaxCut} of a graph of 32 nodes on a publicly available device of only 5 qubits.   This is a $40\%$ increase in graph size compared to state-of-the-art experiments with QAOA on gate-based quantum computers such as the Google Sycamore \cite{harrigan2021quantum} \footnote{The reader should be made aware of the state-of-the-art QAOA experiments with 40 trapped ion qubits \cite{pagano2020quantum} However, here the focus was on finding the ground state of the linear Ising model rather than optimizing a partition of a realistic graph.} . The algorithm presented here opens perspective for immediate quantum speedup with contemporary quantum processors, given that the quantum hardware community is still some years away from producing processors with hundreds of qubits required for quantum speedup with QAOA \cite{guerreschi2019qaoa}. Furthermore, a graph of 128 nodes is partitioned on contemporary simulators of quantum computers. With this methodology, simulators of quantum computers become a powerful tool for graph partitioning, being able to tackle graphs of hundreds of nodes without dividing the problem into smaller subgraphs, such as the work done in \cite{li2021large} or focusing on correlation between pairs of classical independent variables such as work done in Ref. \cite{tan2021qubit}. 

The work presented here maps a \textit{MaxCut} cost function to multi-modal multidimensional cost function which can be evaluated on Quantum Processing Units (QPUs). In the subsequent step it relies on the powerful class of global optimizers called ''genetic optimizers" which are proven to excel in finding minima of multi-dimensional multi-modal black-box cost functions \cite{hoffmeister1990genetic,glibovets2013review,casas2015genetic}. 

\section{Methodology} 

\subsection{Variable reduction} 
In this subsection a variable reduction technique compatible with the \textit{  MaxCut} problem is going to be presented. This is done in order to make the problem amendable to classical optimizers which cannot easily treat multi-dimensional data such as for example the surrogate model EGO optimizer \cite{SMT2019}. Also, as shown in Supplementary Material Fig. S2, variable reduction, yields better results for lower graph densities and lower result variance for  all graph densities. Due to the fact that the binary optimization problem of finding a \textit{MaxCut} is NP-hard, a first approach to approximately solve the problem would be to linearly relax the problem. Meaning that instead of assuming that binary optimization variables in the \textit{MaxCut} problem are integers $0$ or $1$, one assumes that they are continuous variables $[0,1]$. In the field of semi-definite programming \cite{goemans1995improved} a different, more efficient approach is taken, binary variables are substituted with vectors. Such classical method of approximately solving the \textit{MaxCut} problem is state-of-the-art and has a maximum possible performance guarantee of $\alpha=0.87856$ as proved by Goemans and Williamson and can be performed in polynomial time \cite{goemans1995improved}.

\begin{figure}[b!]
	\centering
	\includegraphics[width=0.45\textwidth]{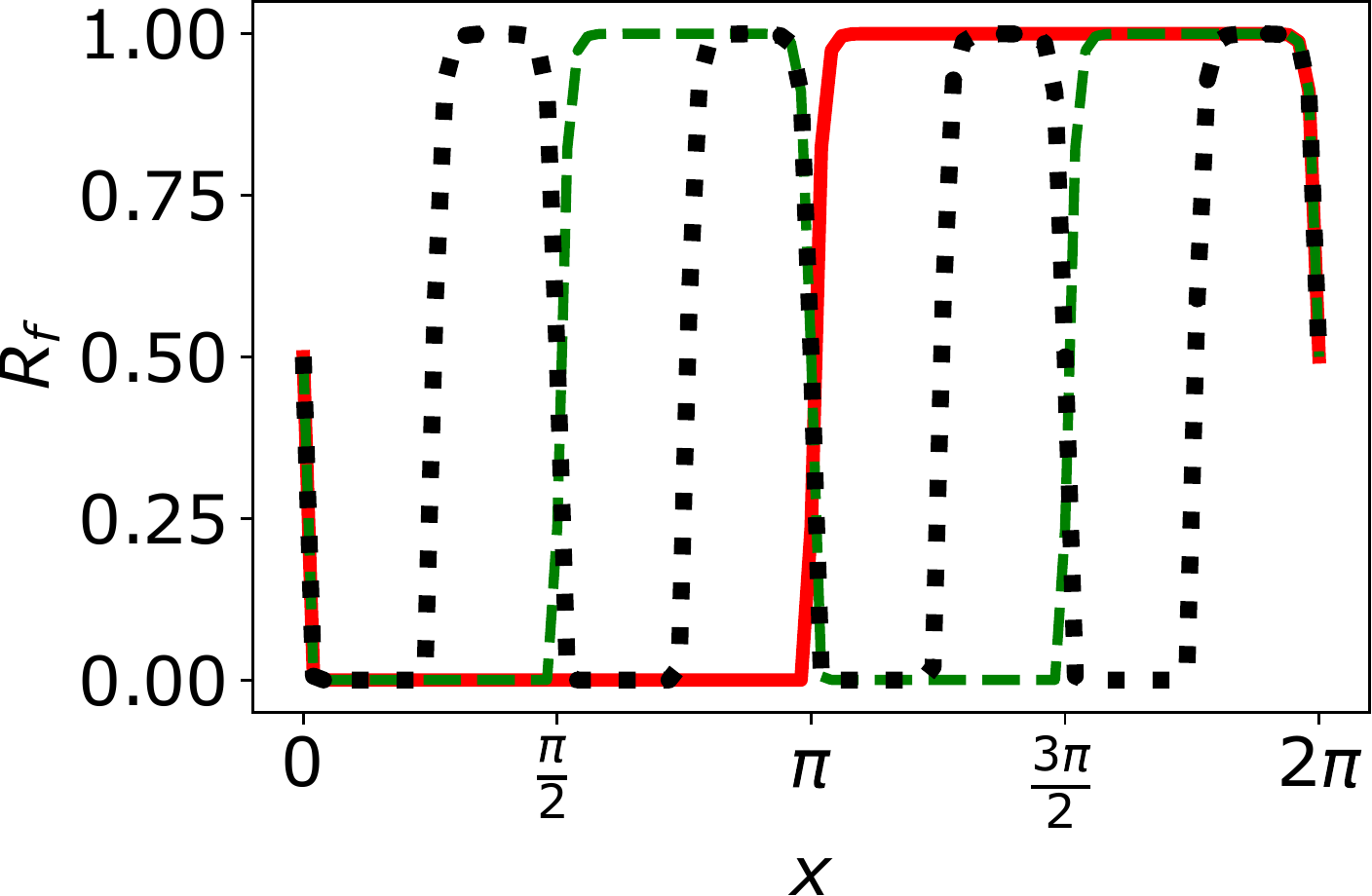}
	\caption{Eq. (\ref{eq:eqRef}) for $q=0$ red, $q=1$ dashed green and $q=2$ dotted black for $m=4$.}
	\label{fig:Fig2}
\end{figure}

Here an alternative approach compatible with quantum computing is presented. Let me now introduce a continuous, differentiable function of the following form
\begin{equation}\label{eq:eqRef}
R_{f}(x,q,m)=\exp{\left(-\exp{\left(2^{m-q}\sin(2^q x+x_0(q,m))\right)}\right)},
\end{equation}
where $x_0(q,m)=\arcsin\left(\ln(-\ln(0.5))/2^{m-q}\right)$ and the integer $m\ge|V|$ and ${0\le q\le |V|-2}$. $R_f$ is defined in such a way to be $n$-differentiable and to have a minimum (maximum) at $0(1)$ and rapidly changing multi-oscillatory behavior in between extremas. It should be noted that this definition is not by any means unique. Assume a graph where $|V|\gg 1$, consequently $m$ is a large number. $R_f(x,0,m)$ is such a function which is mostly $0$ in the region of $0 \le x \le \pi$ and rapidly changes to $1$ in the region $\pi<x\le 2\pi$ (see red line in Figure \ref{fig:Fig2}). I can therefore substitute any binary variable $\theta_1$ taking values 0 or 1 with $\theta_1\rightarrow R_f(x,0,m)$. The second function $R_f(x,1,m)=0$ for $0\le x\le \pi/2$ and $\pi \le x\le 3\pi/2$ and $R_f(x,1,m)=1$ for $\pi/2< x< \pi$ and $3\pi/2 < x <2\pi$. Therefore, I substitute the second binary variable $\theta_2\rightarrow R_f(x,1,m)$ (see green dashed line in Figure \ref{fig:Fig2}). By substituting all $\theta_1,...,\theta_{|V|-1}$ variables with $R_f(x,0,m),...,R_f(x,|V|-2,m)$ I have mapped the $|V|-1$ dimensional binary optimization problem to one-dimensional multi-modal continual variable optimization problem. The role of $x_0(q,m)$ is to center all $Rf$'s to 0.5 at $x=0$, consequently leading to a more regular optimization landscape (avoiding the potential splitting of minima around $0.0$ and $1.0$). It is derived by setting $R_f(0,q,m)=0.5$ for an arbitrary $x_0(q,m)$. 

\subsection{The algorithm} 
A Laplacian of a graph $L(G)$, where $G=(V,E)$ is a $|V|\times|V|$ matrix with $|V|$ positive terms on the diagonal and $2|E|$ off-diagonal terms. The $i$th diagonal term of the graph Laplacian corresponds to the number of connections the node $i$ has with remaining nodes in the graph and the $ij$th off-diagonal term of the matrix is the negative weight between the $i$th and the $j$th node. 

Similarly to a real-valued Hamiltonian in quantum mechanics, the graph Laplacian is symmetric, furthermore it has a spectrum (eigenvalue range) between $0$ and its largest eigenvalue. Now, I introduce a partition vector $\mathcal{V}$ of length $|V|$ with $i$th term equaling $+1$ if $i$th vertex of $G$ belongs to the first sub-graph in the graph partition and $i$th term $-1$ if the $i$th node belongs to the second sub-graph in the graph partition. Then, the number of cuts in the graph bi-partition is $N_{\rm cuts}=\mathcal{V}^T L \mathcal{V}/4$\cite{pothen1990partitioning}, and this formula is a central piece of the algorithm presented here. By finding the vector $\mathcal{V}$ which maximizes $N_{\rm cuts}$ a \textit{MaxCut} of the graph is found. Vector $\mathcal{V}$ has $2^{|V|}$ possible values and there is no known algorithm which can exactly find $\mathcal{V}$ which maximizes $N_{\rm cuts}$ with a computational complexity which is a polynomial of $|V|$.

Now I present the structure of the algorithm for a pre-selection of $r$ optimization variables $x_1...x_r$. \\
1.  Trivial unconnected vertices are added to the graph so it has a dimension which is a power of two.\\
2. The graph Laplacian $L(G)$ is represented as a sum of tensor products of unitary matrices, and denoted as $\mathcal{L}(G)$ in such form. The decomposition is a prerequisite for measuring the expectation value of the graph Laplacian on a QPU (step 5.) and also follows the logic of the implementation in  IBMQ's Qiskit.\\
3. If a graph has $|V|$ nodes a Hadamard gate is applied to $\lceil\log_2|V|\rceil$ qubits. This operation is denoted with $H_{\rm G}$.\\
4. A variational ansatz in a form of diagonal gate is applied
\begin{widetext}
\begin{equation}
U=\rm{diag}\Big(\exp{\Bigl(i\pi R_f(x_1,0,m_r)\Bigr)},..\exp{\Bigr(i\pi R_f(x_1,2^n/r,m_r\Bigl)},...\exp{\Bigl(i\pi R_f(x_r,0,m_r)\Bigr)},..\exp{\Bigl(i\pi R_f(x_r,2^n/r,m_r)\Bigr)},1,...1\Big),
\end{equation}
\end{widetext}
where, the number of variables $r$ is between $1\le r\le 2^n$ and $r \text{ mod }2=0$, and $m_r \ge 2^n/r+2$.\\
5. The number of cuts is calculated as
\begin{equation}\label{eq:NCuts}
N_{\rm cuts}=2^{n-2}\langle0|H_{\rm G} U^{\dagger} \mathcal{L}(G)UH_{\rm G} |0\rangle.
\end{equation}
6. Variational parameter $x$ is adjusted with a classical optimizer and steps 3-5 are repeated until a maximum is reached.

The algorithm presented here maps the \textit{MaxCut} problem of a graph $G=(V,E)$ comprising of $|V|$ vertices and $|E|$ edges to a problem of $|V|$ energy levels coupled with $|E|$ coupling terms described by a Hamiltonian $\mathcal{L}(G)$. The weight between the nodes $w_{ij}$ becomes a coupling strength between energy levels $i$ and $j$. Energy levels $i$ and $j$ are residing at an energy equal to the connectivity of the node $i$ and $j$ respectively.

\begin{figure}
	\centering
	\begin{overpic}[width=0.5\textwidth]{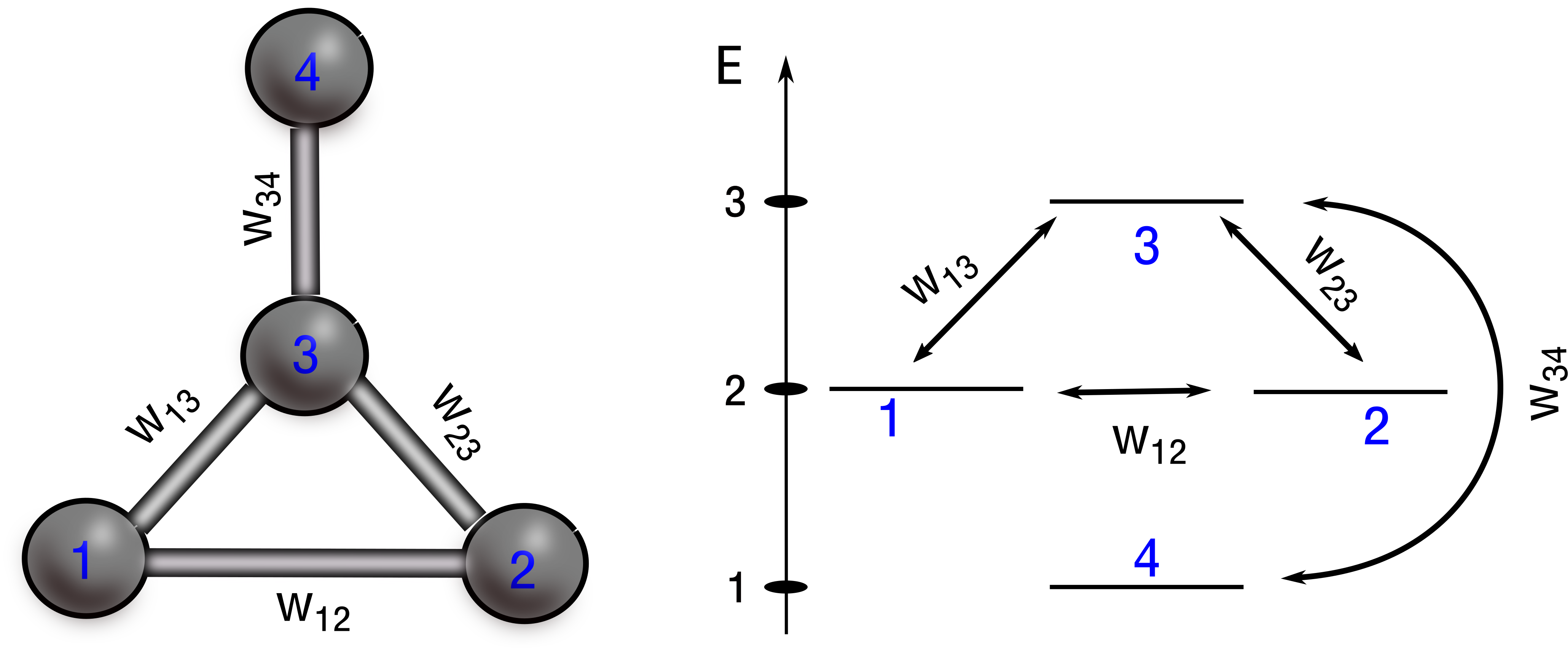}
		\put(-5,40){(a)}
		\put(35,40){(b)}						
	\end{overpic}	
	\caption{(a) A simple graph where $w_{ij}$ are the off diagonal elements of a graph Laplacian. (b) The mapping of the graph to a set of coupled energy levels.}
	\label{fig:Fig1}
\end{figure}

In Figure \ref{fig:Fig1} I represent a simple example of a graph with four nodes. For instance the node $1$ is connected with two other nodes (Figure \ref{fig:Fig1} (a)). Therefore, the energy level $1$ lies at an energy $E=2$ in Figure \ref{fig:Fig1} (b). Node $1$ is connected with nodes $2$ and $3$ so the level $1$ is coupled with levels $2$ and $3$ in the energy scheme. Such logic applies for all nodes of any graph. The algorithm searches for a unitary transformation of the Hamiltonian which maximizes the number of cuts.

\subsection{Circuit depth, computational complexity and quantum advantage} 
Given a graph with $|V|$ nodes $n=\lceil\log_2\left(|V|\right)\rceil$ qubits are required to implement the algorithm. The multi-control multi-target qubit gate on $n$ qubits required to realize the diagonal gate $U$ in Eq. (\ref{eq:NCuts}) can be straightforwardly realized with Grey codes \cite{PhysRevLett.93.130502} or in the context of follow up work \cite{shende2006synthesis} at a cost of $(23/48) \times 4^n- (3/2) \times 2^n+ 4/3$ CNOT gates. However, exploiting the fact that $U$ is a diagonal gate and following on the works of \cite{shende2006synthesis} and \cite{malvetti2021quantum} it can be realized with $2^n-2$ CNOT gates. This means that the ansatz is implemented with less than $ 2|V|$ two qubit CNOT gates which is in stark contrast with the QAOA ansatz which requires $p |E|$ two qubit gates, where $p$ is the depth parameter. Given that $|E|\gg|V|$ the algorithm presented here is much more efficient in the number of two qubit gates as compared to even the lowest depth $p=1$ QAOA. 

The algorithm presented here is a heuristic, meaning that its depth is case dependent. However, the quantum implementation of the heuristic can be compared with its classical counter-part for every step of the evaluation. A classical computing variant of Eq. (\ref{eq:NCuts}) is a vector-matrix-vector multiplication. For a $|V|\times |V|$ matrix the computational complexity of such an evaluation is $O(|V|^2)$. 

 In the course of this study the graph Laplacian is decomposed into sum of Pauli matrices with Pennylanne's \textit{qml.Hermitian} function \cite{bergholm2018pennylane}, afterwards a conversion to \textit{Qiskit} was performed. Mathematically, the Pauli basis decomposition of a general Hermitian matrix $H$ is conducted in the following fashion

\begin{equation}\label{eq:dec}
H=\frac{1}{4}\sum_{i=1}^{4^n< 2|V|^2} {\rm Tr} (P_i\, H ) P_i,
\end{equation}
where $P_i$ are all possible combinations of tensor products of the Pauli group composed of $\{I,X,Y,Z\}$ acting on $n$ qubits. It should be noted that the process of decomposing a Hermitian matix into sum of Paulis (see Eq. \ref{eq:dec}) is completed in $O\left(|V|^2\right)$: up to $|V|^2$ summands, $|V|^3$ for the matrix-matrix product of $P_i$ and  $H$ and $|V|$ for computing the trace. For example, for $n=2$ qubits (up to $|V|=4$ vertex graph) $4^n=|V|^2=16$ possible products exist

\begin{align}
\begin{split}
P_i  &\in \{I I,IX,I Y,IZ,XI,XX,XY,XZ,
Y I,Y X,YY,\\&YZ,YI, ZX,ZY,ZZ\}.
\end{split}
\end{align}
It should be noted that all tensor products which are complex in nature (involving odd numbers of $Y$ Pauli matrices per tensor product) yield a zero trace due to the real nature of the graph Laplacian in our study. Furthermore, in our case the graph Laplacian is a real Hermitian matrix and is thus symmetric, so only the decomposition for lower or upper triangular parts could be performed. Finally, such decompositions are easily parallelized on HPC architectures as all computations are independent one of another. 

On a quantum computer the computational complexity of evaluating Eq. (\ref{eq:NCuts}) is $O(|V|^3)$ (one power of $|V|$ coming from the ansatz and up to $|V|^2$ summands yielding $\mathcal{L}(G)$). This number although polynomial in $|V|$ can still be quite high for large graphs and thus would require an estimation of the expectation value of the large number of summands on a quantum computer. However, simulating a $d-$sparse Hamiltonian ($d-$ regular graph) is done in maximally $O(d^2(d+\log^*n))$ queries with the so-called star-decomposition of the Hamiltonian \cite{berry2007efficient,childs2010simulating}, and efficiently simulating a Hamiltonian is equivalent to calculating an expectation value of a Hamiltonian \cite{PhysRevA.75.012328}. So for $d-$regular graphs every step in the heuristic algorithm executes in $O(|V|d^2(d+\log^*\lceil\log{|V|\rceil}))$ which is smaller than the classical $O(|V|^2)$ for small $d${\tiny }. Further reductions of the number of measurements could be conducted by the command group$\_$paulis$=$True in Pauli expectation class of qiskit. This allows grouping of Pauli strings which commute into same measurements. Also, with larger processors – the tasks could be parallelized – certain qubits just estimating certain sets of Pauli stings.

In Table \ref{tab:tab} I summarize the main differences between the algorithm presented here and the QAOA. The diagonal gates required to preform the algorithm presented here requires an all-in-all connectivity between qubits for optimal performance. On the other hand, QAOA performs best when the qubits are connected in the same way as the nodes of the graph \cite{harrigan2021quantum}.

As of 2021 the second most powerful supercomputer in the world is Fugaku \cite{top500list} with $158,967$ nodes each having $32$ Gb of RAM. In total, this supercomputer can store $158,967 \times 32$  Gigabytes of data, equaling to it being able to handle $5\cdot 10^{15}$ bytes of data. Commonly one requires $8$ bytes to store a real  number \cite{goldberg1991every,overton2001numerical}, indicating that $0.64 \cdot 10^{15}$ double precision numbers can be stored in the RAM memory in such a device. Given that a graph Laplacian is a square matrix, the dimensions of a weighted graph Laplacian which such a supercomputer could handle is $\sqrt{0.64\cdot10^{15}}\times \sqrt{0.64\cdot10^{15}}=25.2\cdot10^{6}\times25.2 \cdot 10^6$. On the other hand side, a $25$ qubit device (such as those already available at IBM and Google) could handle a $2^{25} \times 2^{25}=33.5\cdot10^6 \times 33.5\cdot 10^6$ graph Laplacian. Of course, noise would be a limiting factor in handling such sizes of optimization problems in the pre-error correction era. However, next year is going to be the year in which large scale error mitigation is going to be implemented on IBMQ systems \cite{IBMQ_roadmap}. This will increase the depth of circuits which could be executed on IBM QPUs to a larger level which is difficult to estimate at this point. It should also be noted that the way of calculating the number of cuts of a graph on a quantum computer as given by Eq. (\ref{eq:NCuts}) and can also be applicable to a plethora of algorithms handling different aspects of graph cuts, not only \textit{MaxCut} with the genetic optimizer as done in this study.

\begin{table}[h!]
	\begin{center}
		\begin{tabular}{ |c|c|c|c|} 
			
			\hline
			
			& QAOA & New alg.  & New alg. $d-$regular gr. \\ 
			\hline			
			\hline			
			Ansatz & $O(p\left(|E|+|V|\right))$ & $O(|V|)$& $O(|V|)$ \\ 
			\hline			
			Algorithm     & $O(p\left(|E|+|V|\right))$ & $O(|V|^3)$& $O(|V|d^2(d+\log^*n))$ \\ 
			\hline	
			Qubit nr. & $|V|$& $n$& $n$\\ 
			\hline	
			Connectivity  & graph inspired & all-in-all& all-in-all\\
			\hline
		\end{tabular}
	\end{center}
	\caption{A table summarizing the difference between the approach presented here and QAOA. Complexities are given for one evaluation step and  $n=\lceil\log_2{|V|} \rceil$. \label{tab:tab}}
\end{table}	

\section{Results}

\begin{figure}[t!]
	\includegraphics[width=0.5\textwidth]{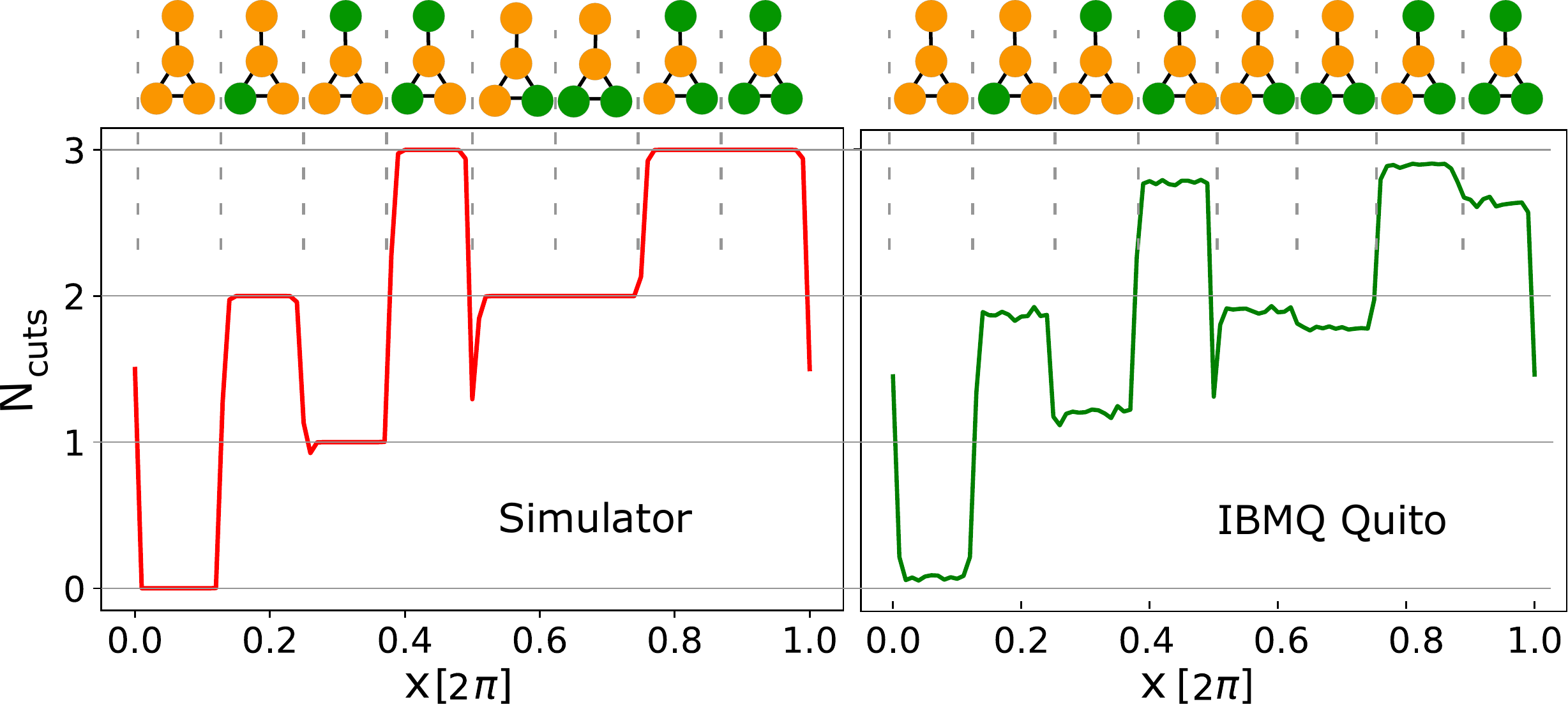}
		\caption{The number of cuts of a 4-node graph obtained with 2 qubits on a simulator of quantum computers (left) and an actual quantum device (right). \label{fig:CostF}}
\end{figure}

In Figure \ref{fig:CostF} I compare the output of a simulator with the output of publicly available IBMQ Santiago. Although pure dephasing, shot noise and relaxation may distort the optimization landscape, maxima are clearly noticed although equal local maxima become unequal. The $N_{\rm cuts}$ is estimated for 100 equidistant values of $x$.

I further present results obtained by benchmarking randomly generated 3-regular graphs of 32 nodes on actual quantum computers and simulators and randomly generated 3-regular graphs of 128 nodes on a simulator of quantum computers (Qiskit). Intensive testing showed that the algorithm performs best when the number of variables is kept at 8 or 16 for graphs of the size 32-128 nodes. Intensive numerical testing also showed that a genetic optimizer is best suited for finding the maximum of the function - not too surprising as genetic optimizer is indeed best used for multi-modal cost functions. On top of the genetic algorithm, a number of classical optimizers were tried (COBYLA, Neldear-Mead, Basin-Hopping, Particle Swarm, EGO). Further details about the setting of the genetic optimizer can be found in the Supplementary Material S1.

\begin{figure*}
	\centering	
	\begin{minipage}{0.33\textwidth}
		\begin{overpic}[width=1.0\textwidth]{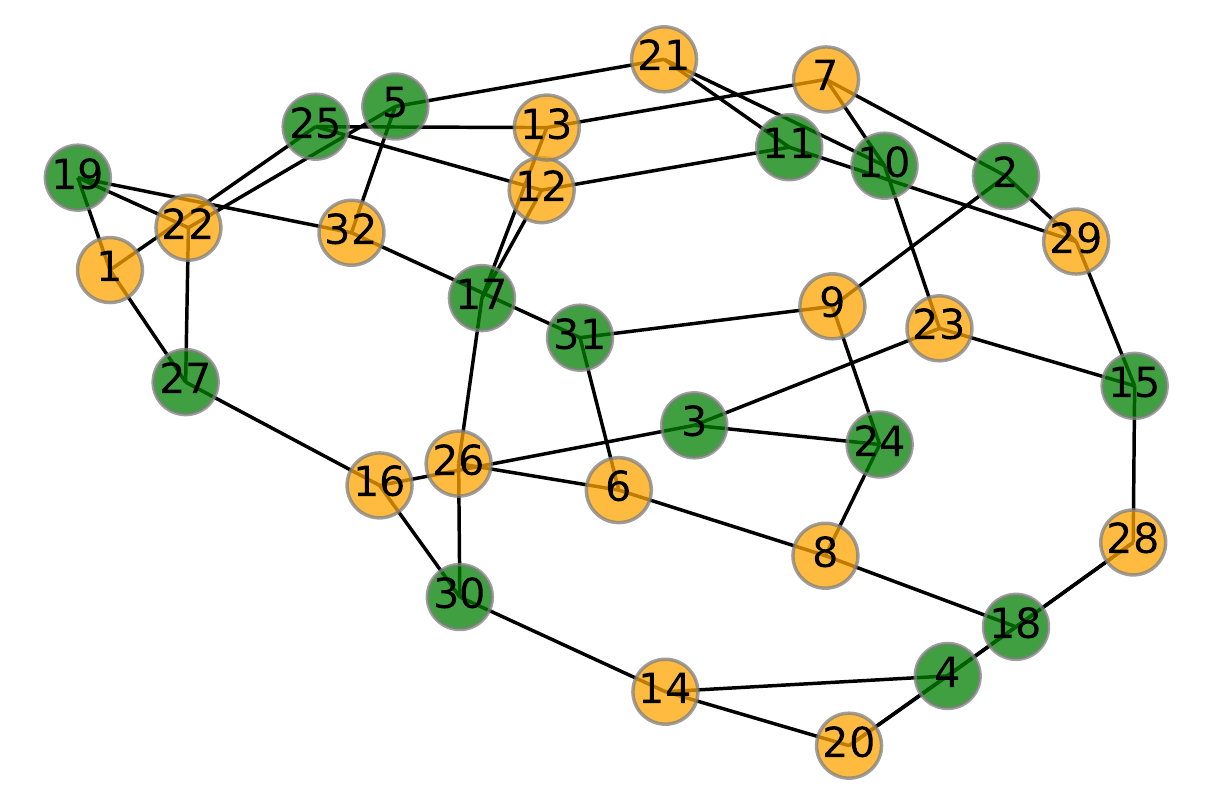}
			\put(0,70){(a) GW \textit{MaxCut} $=43$}
		\end{overpic}
	\end{minipage}\begin{minipage}{0.33\textwidth}
		\begin{overpic}[width=1.0\textwidth]{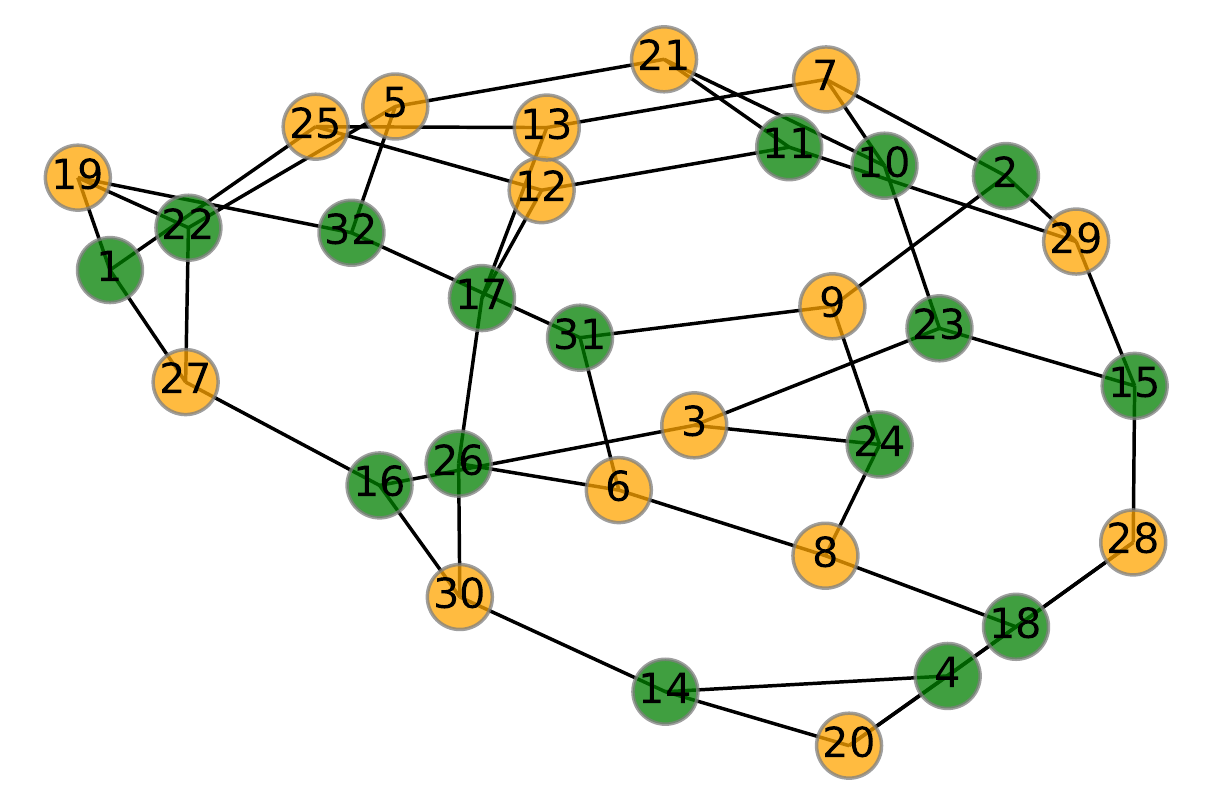}
			\put(0,70){(b) Qiskit \textit{MaxCut} $=38$}
		\end{overpic}
	\end{minipage}\begin{minipage}{0.33\textwidth}
		\begin{overpic}[width=1.0\textwidth]{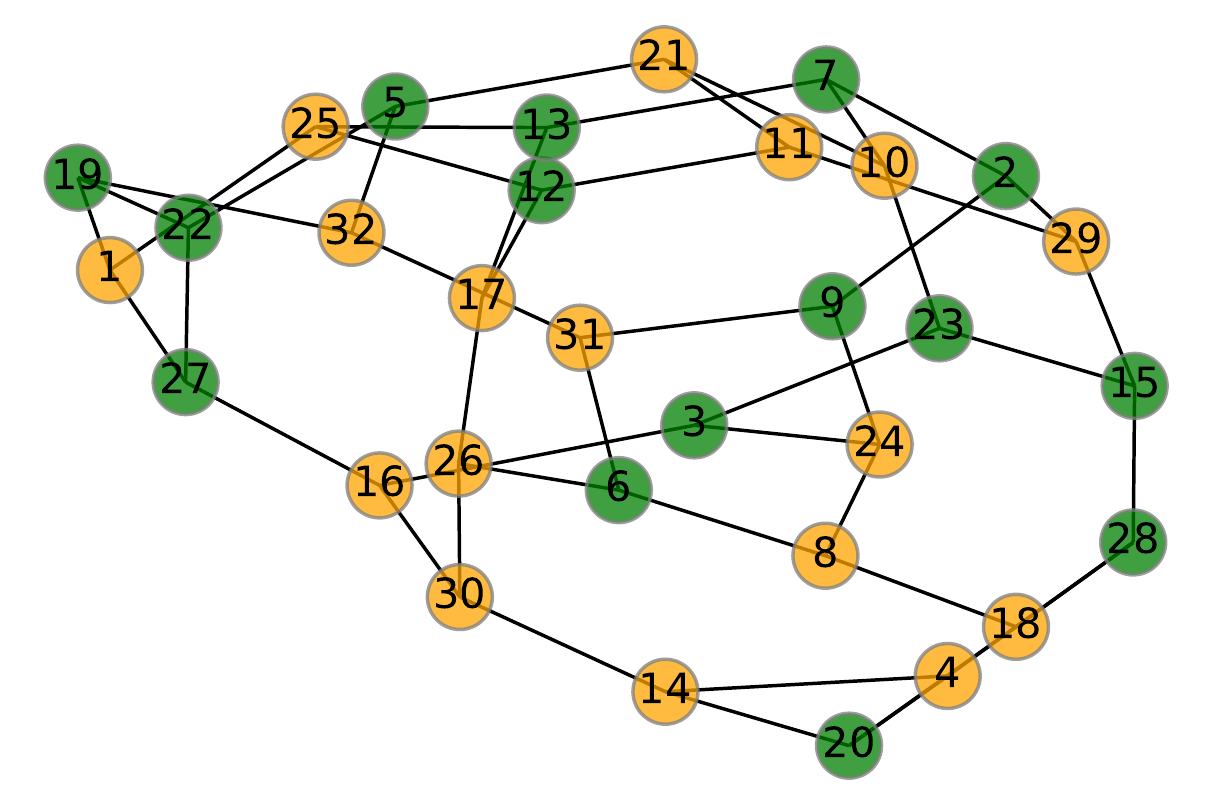}
			\put(0,70){(c) IBMQ Quito \textit{MaxCut} $=27$}
		\end{overpic}
	\end{minipage}
	
	\vspace{1cm}
	
	\begin{minipage}{0.33\textwidth}
		\begin{overpic}[width=1.0\textwidth]{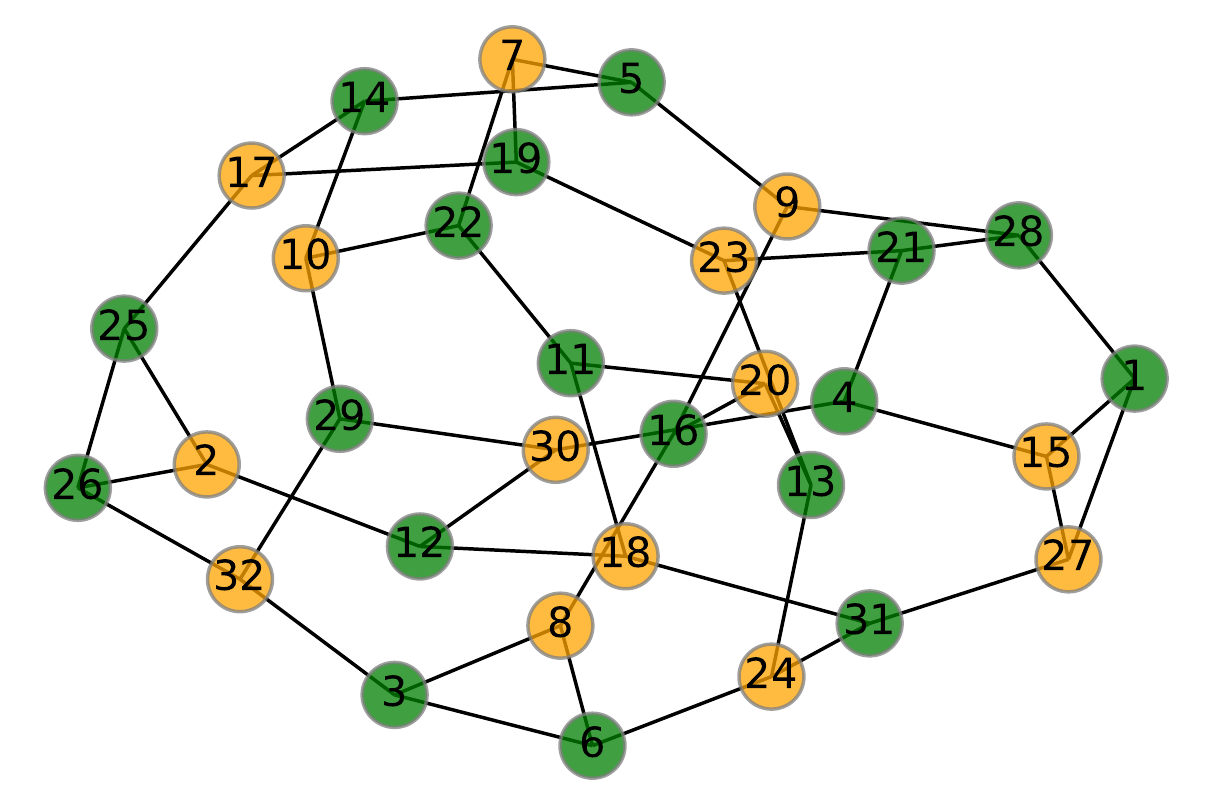}
			\put(0,70){(d) GW \textit{MaxCut} $=40$}
		\end{overpic}
	\end{minipage}\begin{minipage}{0.33\textwidth}
		\begin{overpic}[width=1.0\textwidth]{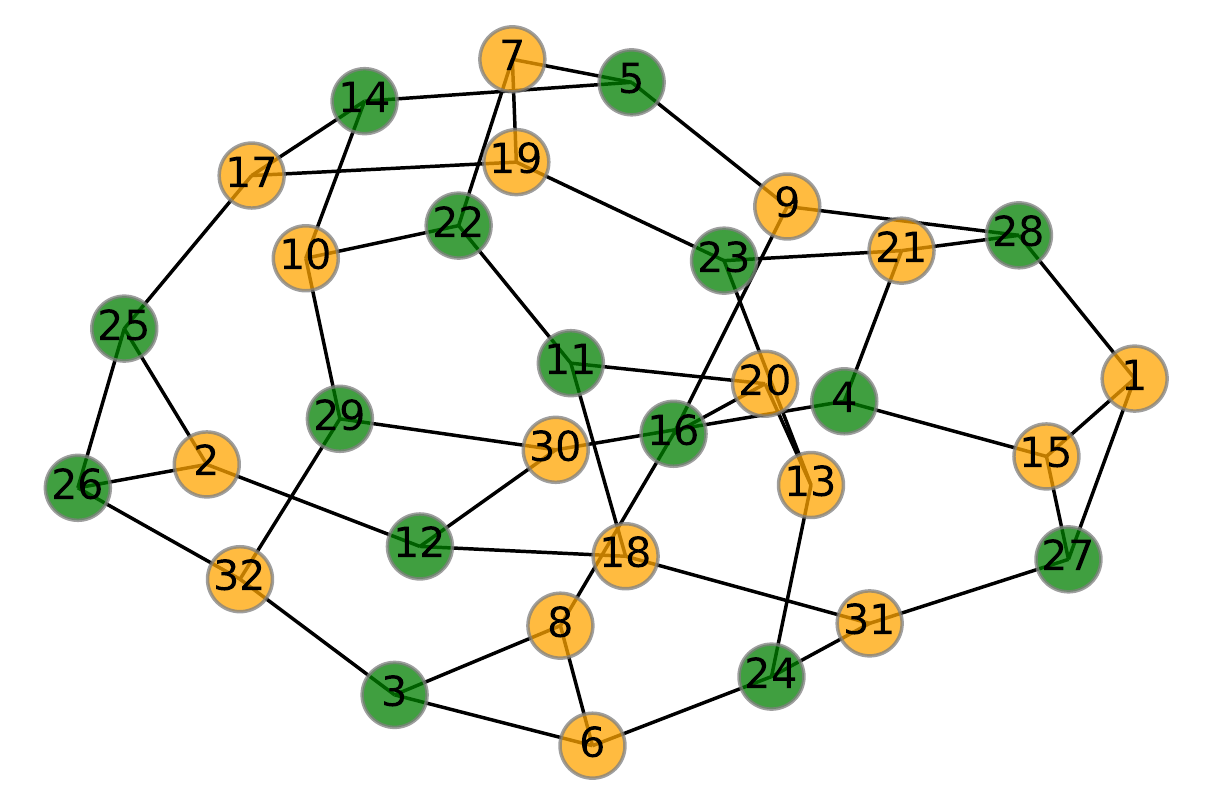}
			\put(0,70){(e) Qiskit \textit{MaxCut} $=39$}
		\end{overpic}
	\end{minipage}\begin{minipage}{0.33\textwidth}
		\begin{overpic}[width=1.0\textwidth]{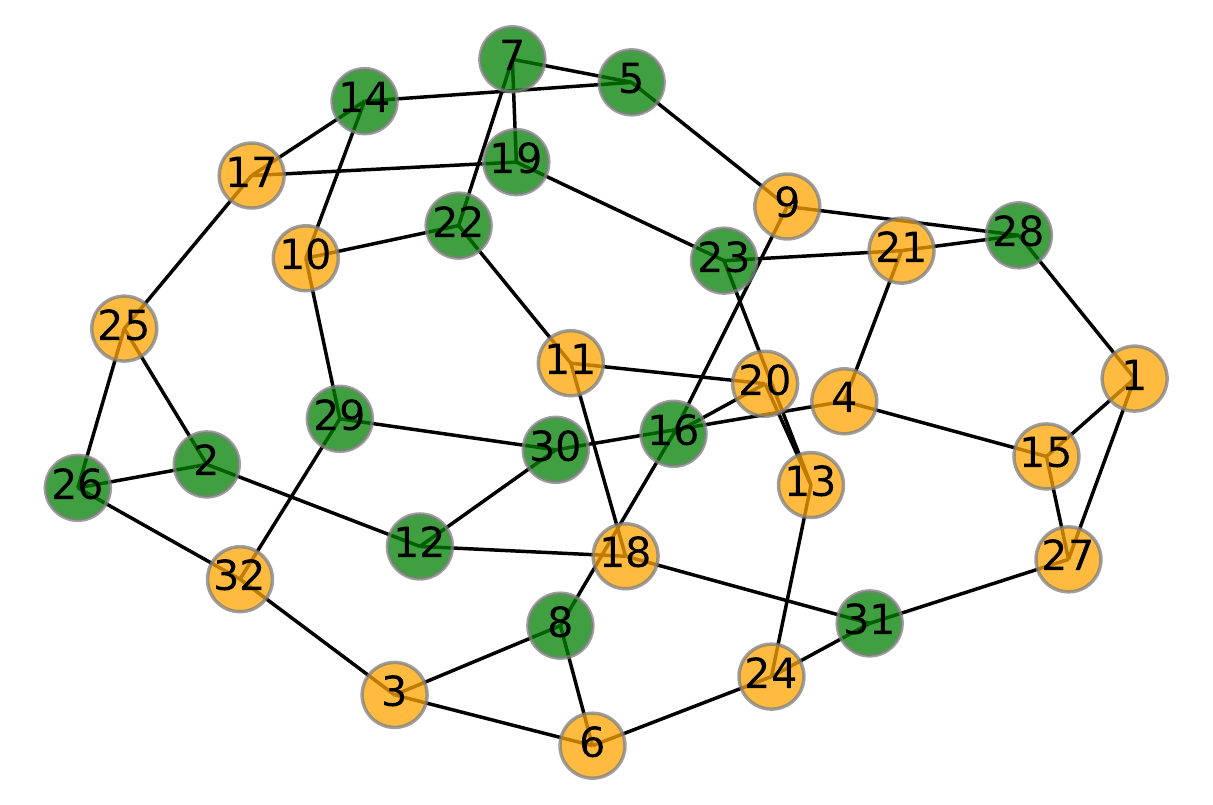}
			\put(0,70){(f) IBMQ Santiago \textit{MaxCut} $=25$}
		\end{overpic}
	\end{minipage}
	\caption{Randomly generated 3-regular graphs of 32 nodes. Nodes belonging to different partitions are marked in green and orange respectively and the \textit{MaxCut} value is written on top of the graph. Graphs are randomly initialized by Python's networkx package where seed 20 is used for graphs (a-c) and seed 30 for graphs (e-g). GW stands for Goemans-Williamson. \label{fig:cut1}}
\end{figure*}

For the case of 3-regular graphs of 32 nodes variable reduction is preformed so that the optimization landscape has 8 variables. The \textit{MaxCut} is calculated with 5 qubits. The Goemans-Williamson algorithm (GW) is a classical approximate algorithm for the \textit{MaxCut} problem has a performance guarantee of 0.87856 \cite{goemans1995improved}. I define the approximation ratio of the algorithm presented here with respect to the exact solution as
\begin{equation}
0.87856\,\textit{MaxCut}/\textit{MaxCut}_{\rm GW}\le r \le \textit{MaxCut}/\textit{MaxCut}_{\rm GW},
\end{equation}
 where \textit{MaxCut} is the value obtained with the algorithm presented here and $\textit{MaxCut}_{\rm GW}$ is the value obtained with Goemans-Williamson. For the first graph Figure \ref{fig:cut1} (a-c) the algorithm performed on a simulator of quantum computers yields $0.776 \le r\le 0.884$ and the algorithm executed on IBMQ Quito $0.552 \le r \le0.628$. For the second graph Figure \ref{fig:cut1} (d-f) the algorithm preformed on a simulator of quantum computers yields $0.857 \le r\le 0.975$ and the algorithm executed on IBMQ Quito $0.549 \le r \le0.625$. For both realizations there is a clear difference between actual hardware benchmarks and ideal simulation. I assume that the main reason for this is the distortion of the optimization landscape due to pure dephasing and relaxation. I expect that shot noise contributed less as the expectation value was estimated for the maximally allowed 8192 shots.
 
For the case of 3-regular graphs of 128 nodes variable reduction is performed so that the optimization landscape has 16 variables. The \textit{MaxCut} is calculated with 7 simulated qubits under the assumption of no noise processes. Given that devices larger than 5 qubits are unavailable to the author, for these 3 graphs I stayed in the domain of quantum simulators. For the first graph Tab. \ref{tab:tab2} seed 7 $0.679 \le r \le 0.773$, second graph Tab. \ref{tab:tab2} seed 8 $0.743 \le r\le 0.846$, third graph Tab. \ref{tab:tab2} seed 9 $0.709 \le r\le 0.807$. Values do not converge as nicely as for smaller graphs likely because the genetic algorithm gets trapped in a local minimum with increasing system size. These results are visually represented in Supplementary Material S3.

\begin{table}[h!]
	\begin{center}
		\begin{tabular}{ |c|c|c|} 
			
			\hline
			
			seed & \textit{Maxcut}(GW)& \textit{Maxcut}(New alg.)\\ 
			\hline			
			\hline			
			7 & 172&133 \\ 
			\hline			
			8    & 162 & 137 \\ 
			\hline	
			9 & 166& 134\\ 
			\hline	
		\end{tabular}
	\end{center}
	\caption{A table summarizing the results of benchmarking graphs of 128 nodes on a quantum simulator. A random seed in Python is used to generate graph instances. GW stands for the Goemans-Williamson algorithm.  \label{tab:tab2}}
\end{table}	

A $d$-regular graph with $|V|$ nodes has $d\times|V|/2$ edges \cite{aldous2003graphs}. An average random bi-partition of such a graph is $d\times |V|/4$\cite{motwani1995randomized}, or in the case of 3-regular graphs with 32 nodes the average random bi-partition is 24. So both the quantum hardware results and the simulator results stay above the average random bi-partition value. An average random bi-partition of a 3-regular graph of 128 nodes is $3\times 128/4=96$.
\section{Conclusion}
In conclusion I have presented a novel algorithm for noisy intermediate-scale quantum computers requiring logarithmically less qubits and significantly less quantum gates as compared to the contemporary state-the-of-art algorithm - QAOA. I went through to calculate the \textit{MaxCut} of a randomly generated 3-regular graph of 32 nodes, a $40\%$ increase compared to experiments of state-of-the-art gate-based quantum computers (Google Sycamore). I did so with publicly available IBM hardware, and obtained a lower bound of $54.9\%$ on the solution for actual hardware benchmarks and $77.6\%$ on ideal simulators of quantum computers. Furthermore, I calculated the \textit{MaxCut} of a 3-regular graph of 128 nodes with quantum simulator obtaining a lower bound of $67.9 \%$ on the solution and with no pre-processing of the graph what-so-ever. 

\section{Acknowledgments, funding, contributions, competing interests, data availability}
I acknowledge discussions with Adrien Suau, Vedran Dunjko, Thomas B\"{a}ck and Zaid Allybokus, with the latter especially in the domain of use cases to which the algorithm may be applied.  I am thankful to the EU for funding within the H2020 project $\langle$NE$|$AS$|$QC$\rangle$.  Marko J. Ran\v{c}i\'{c} envisioned, developed and conducted the whole study himself. The work presented here is a part of a broader provisional patent claim ''Method for optimizing a functioning relative to a set of elements and associated computer program product" submission number EP21306155.9 submitted on 26.8.2021. with Marko J. Ran\v{c}i\'{c} and Zaid Allybokus being listed as inventors. The author declares no further competing interests.  Output from quantum computers and codes to support the claim are available at DOI: 10.5281/zenodo.5592834

\bibliographystyle{unsrt}
\bibliography{PaperV_1}
\newpage
\begin{widetext}
\begin{center}
	\textbf{\large Supplementary Material: Noisy intermediate-scale quantum computing algorithm for solving an $n$-vertex MaxCut problem with log($n$) qubits.}
\end{center}

\setcounter{equation}{0}
\setcounter{section}{0}
\setcounter{figure}{0}
\setcounter{table}{0}
\setcounter{page}{1}
\makeatletter
\renewcommand{\theequation}{S\arabic{equation}}
\renewcommand{\thefigure}{S\arabic{figure}}
\renewcommand{\thesection}{S\arabic{section}}

\section{S1 - The classical optimizer}\label{sec:claso}

After intensive numerical testing a genetic optimizer performed the best in converging to the maximal cost function. An open GitHub code was incorporated Ref. \cite{rmsolgi}. A genetic algorithm is an optimizer inspired in evolution \cite{katoch2021review}. To obtain results displayed in Figure 4. (main body of the paper) and Figure \ref{fig:cut2} the optimization settings in Figure \ref{fig:optset} were used. For results displayed in Figure \ref{fig:cut2} a maximum of 200 iterations were set with a population size of 14.

In Fig. \ref{fig:NumVar} I show a classical simulation of the genetic algorithm applied on \textit{MaxCut} problem with varying graph densities and number of variables. "m" denotes the best value of the \textit{MaxCut}, among optimizations with different number of variables. Noise level of 15$\%$ was artificially induced to mimic the behavior of shot noise, pure dephasing and relaxation of a QPU.

The result indicate that for randomly generated graphs of low density, it is beneficial to perform variable reduction, while for dense graph (density$>0.27$, beyond the scope of this work) the genetic optimizer in average performed better without variable reduction (x-symbols). However, the variance between minima and maxima of the found \textit{MaxCut} (shaded areas) was larger without variable reduction (shaded areas) for all instances. It can be concluded that adding multimodalty in the cost function while reducing dimensions stabilizes the variance of the result between different runs.

\begin{figure}[H]
	\centering
	\includegraphics[width=0.45\textwidth]{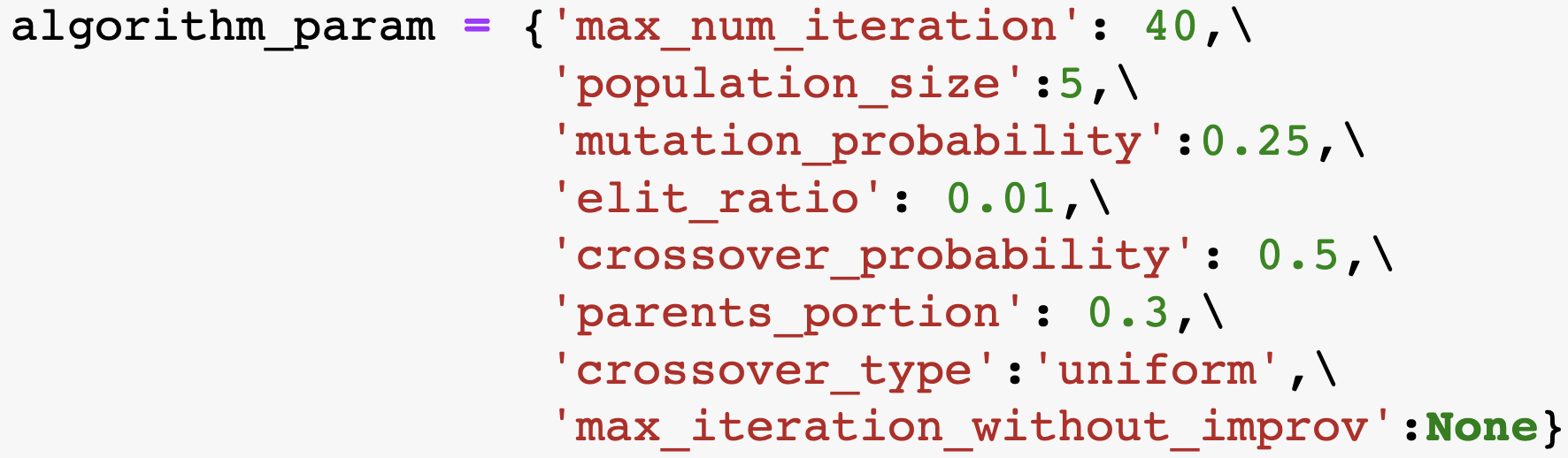}
	\caption{The genetic algorithm optimizer settings used in Figure 4 (main body of the paper). \label{fig:optset}}
\end{figure}

\begin{figure}[H]
	\centering
	\includegraphics[width=0.45\textwidth]{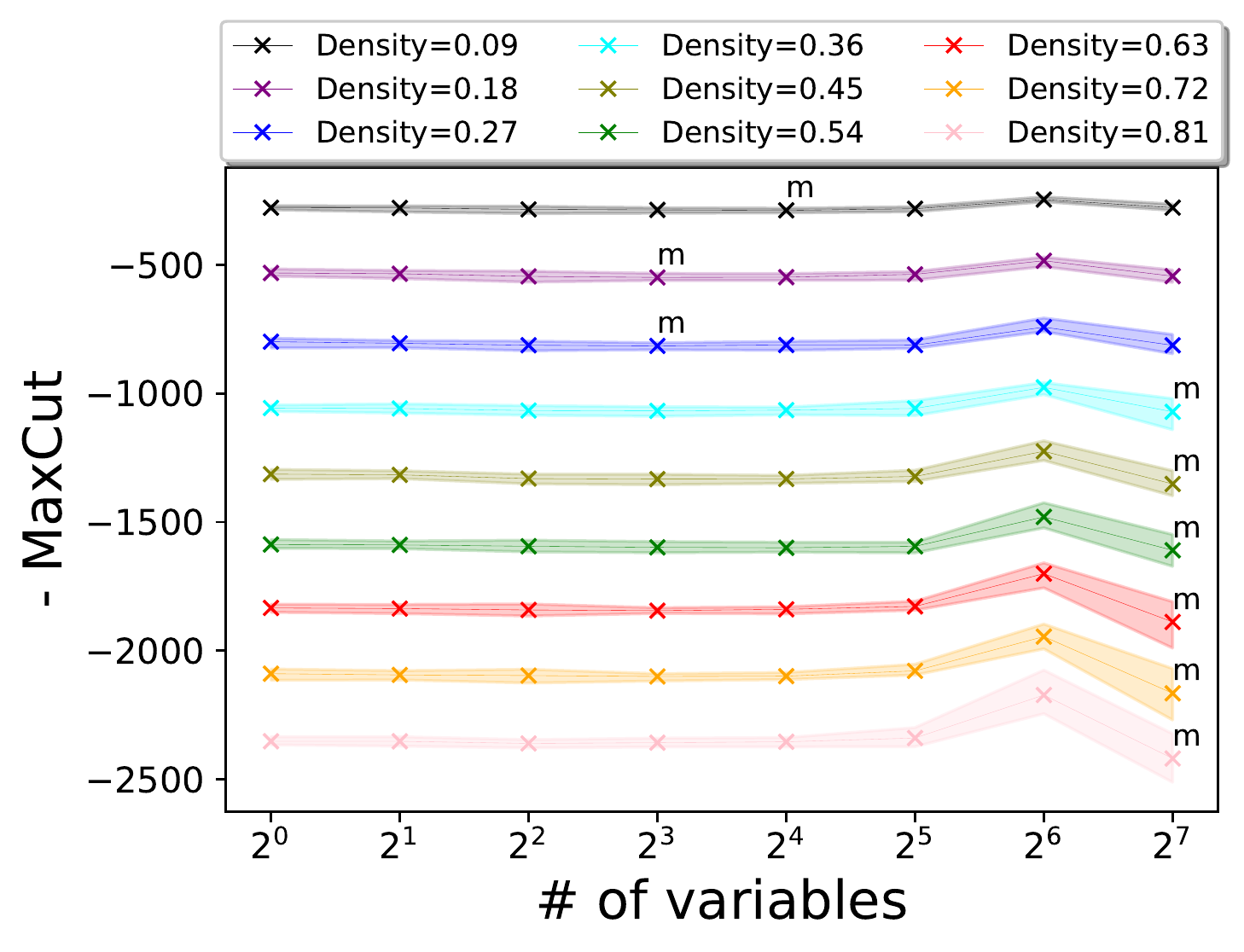}
	\caption{A classical simulation: negative MaxCut as a function of number of optimization variables for random seed 2 of a 128 vertex graph with varying density and 15$\%$ random uncertainty in cost function calculation. Optiimizeer was set like in Fig. \ref{fig:optset}, every point is a average of 20 repetitions. \label{fig:NumVar}}
\end{figure}

In Figure \ref{fig:CostEv} I show how the genetic optimizer improves the cost function as a function of iteration. The value is negative as the optimizer is tuned to search for the minimum of the negative graph Laplacian under a unitary transformation.

\section{S2 - Visually representing the 128 node graph cuts}\label{sec:reprez}
In Fig. \ref{fig:cut2} I visually represent results of Tab. II from the main body of the paper.

\begin{figure}[H]
	\centering	
	\begin{minipage}{0.39\textwidth}
		\begin{overpic}[width=0.9\textwidth]{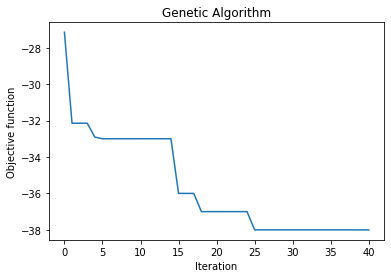}
			\put(0,70){(a) Qiskit, seed 20}
		\end{overpic}
	\end{minipage}\begin{minipage}{0.39\textwidth}
		\begin{overpic}[width=0.9\textwidth]{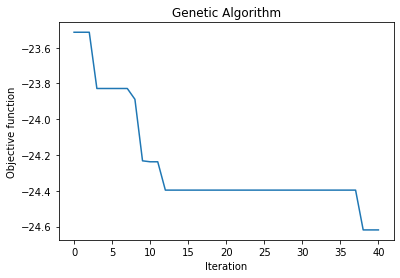}
			\put(0,70){(b) IBMQ Quito, seed 20}
		\end{overpic}
	\end{minipage}
	
	\vspace{0.5cm}
	
	\begin{minipage}{0.39\textwidth}
		\begin{overpic}[width=0.9\textwidth]{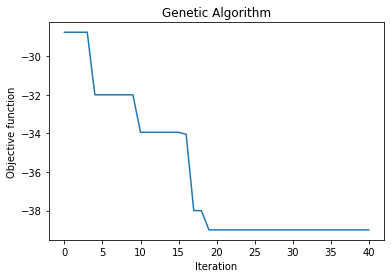}
			\put(0,70){(c) Qiskit, seed 30}
		\end{overpic}
	\end{minipage}\begin{minipage}{0.39\textwidth}
		\begin{overpic}[width=0.9\textwidth]{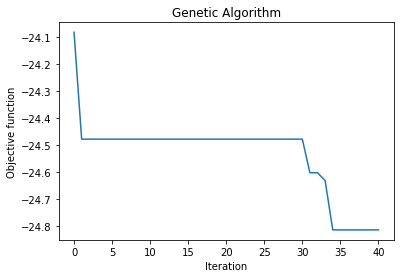}
			\put(0,70){(d) IBMQ Santiago, seed 30}
		\end{overpic}
	\end{minipage}
	\caption{Cut as a function of the number of iterations for Fig. 4 main body of the paper. \label{fig:CostEv}}
\end{figure}

\begin{figure}[H]
	\centering	
	\begin{minipage}{0.46\textwidth}
		\begin{overpic}[width=1.0\textwidth]{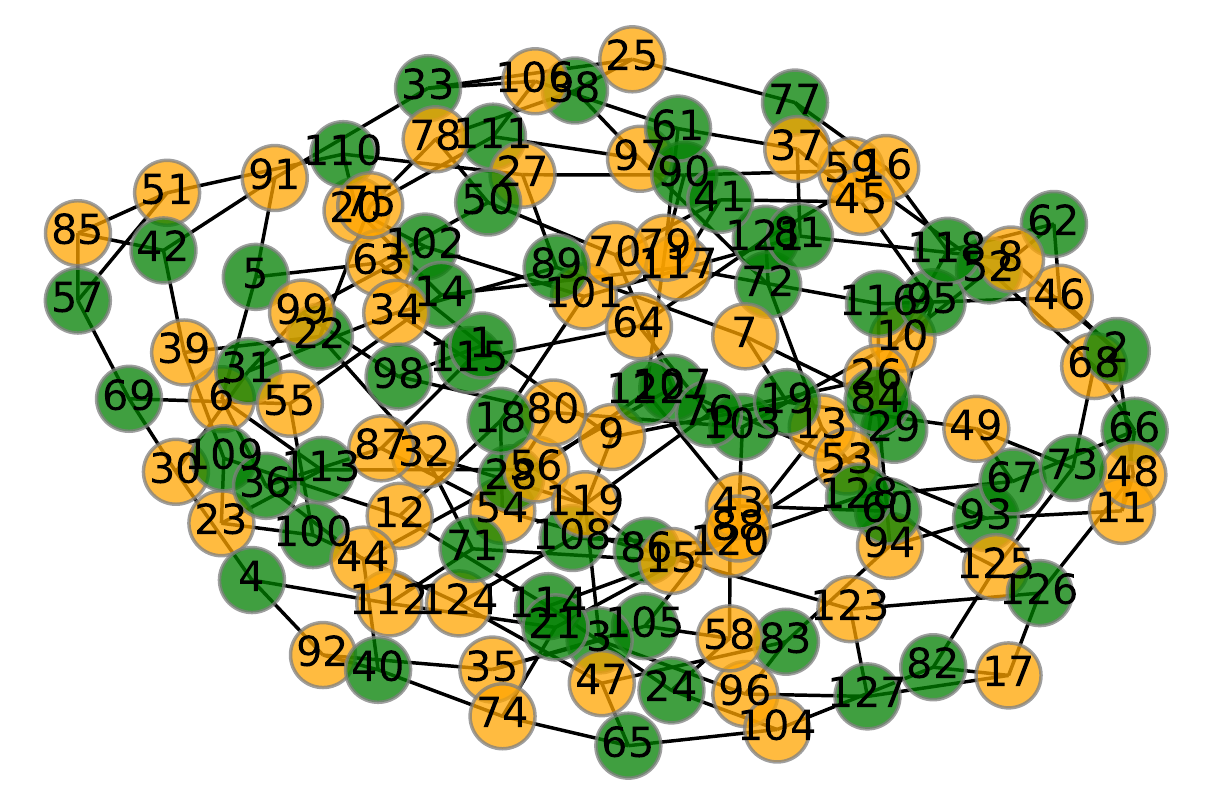}
			\put(0,70){(a) GW \textit{MaxCut} $=172$}
		\end{overpic}
	\end{minipage}\begin{minipage}{0.46\textwidth}
		\begin{overpic}[width=1.0\textwidth]{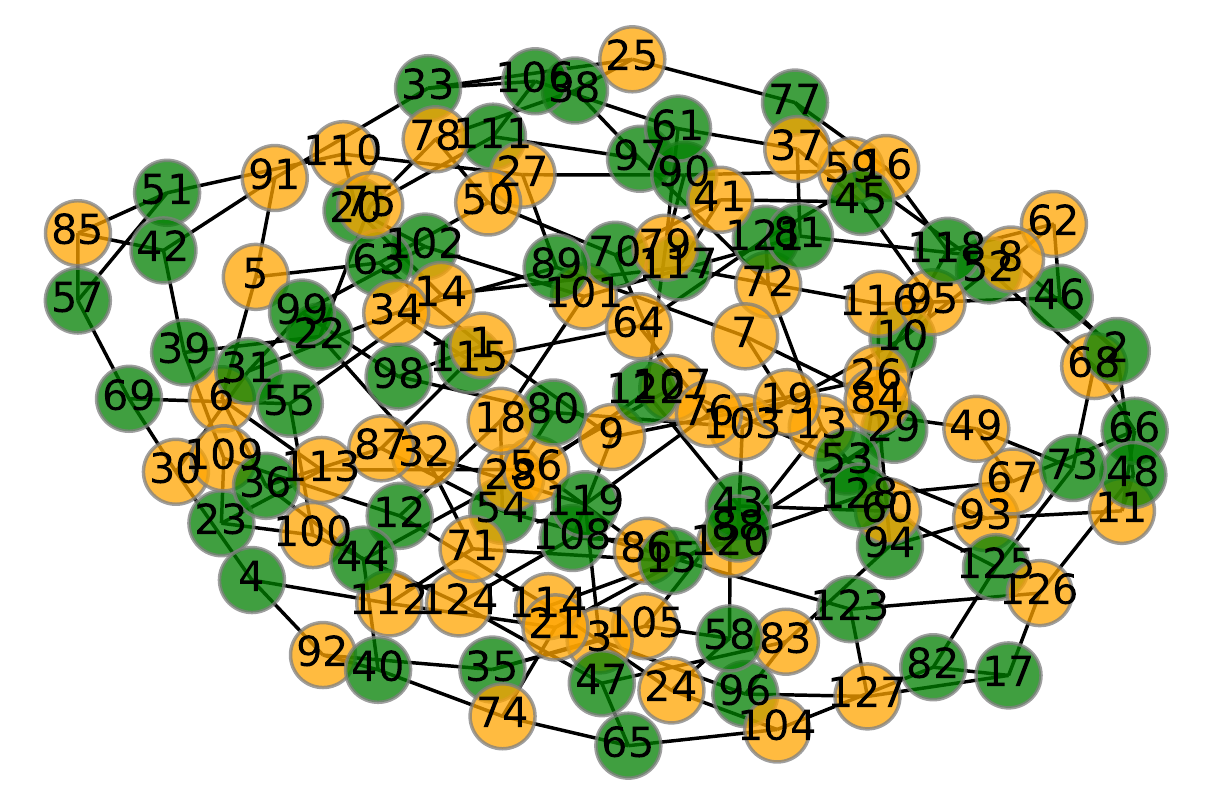}
			\put(0,70){(b) Qiskit \textit{MaxCut} $=133$}
		\end{overpic}
	\end{minipage}
	
	\vspace{1cm}
	
	\begin{minipage}{0.46\textwidth}
		\begin{overpic}[width=1.0\textwidth]{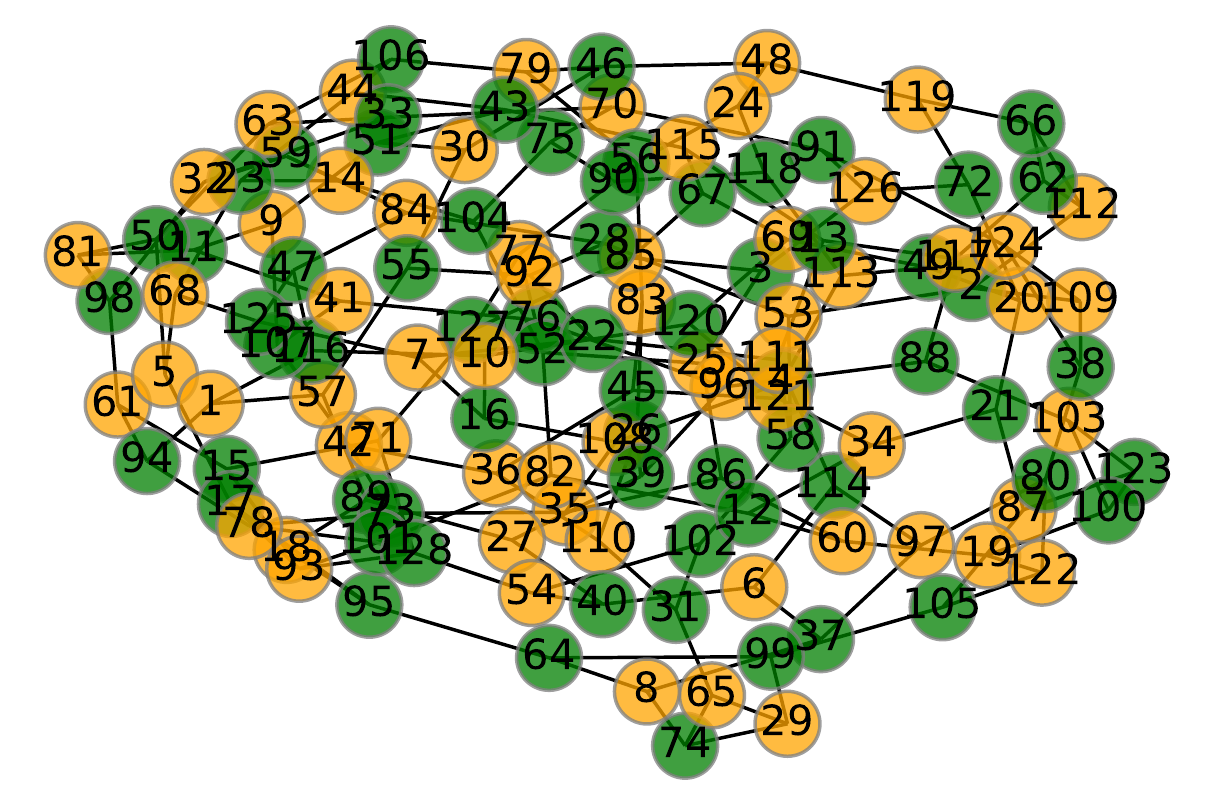}
			\put(0,70){(c) GW \textit{MaxCut} $=162$}
		\end{overpic}
	\end{minipage}\begin{minipage}{0.46\textwidth}
		\begin{overpic}[width=1.0\textwidth]{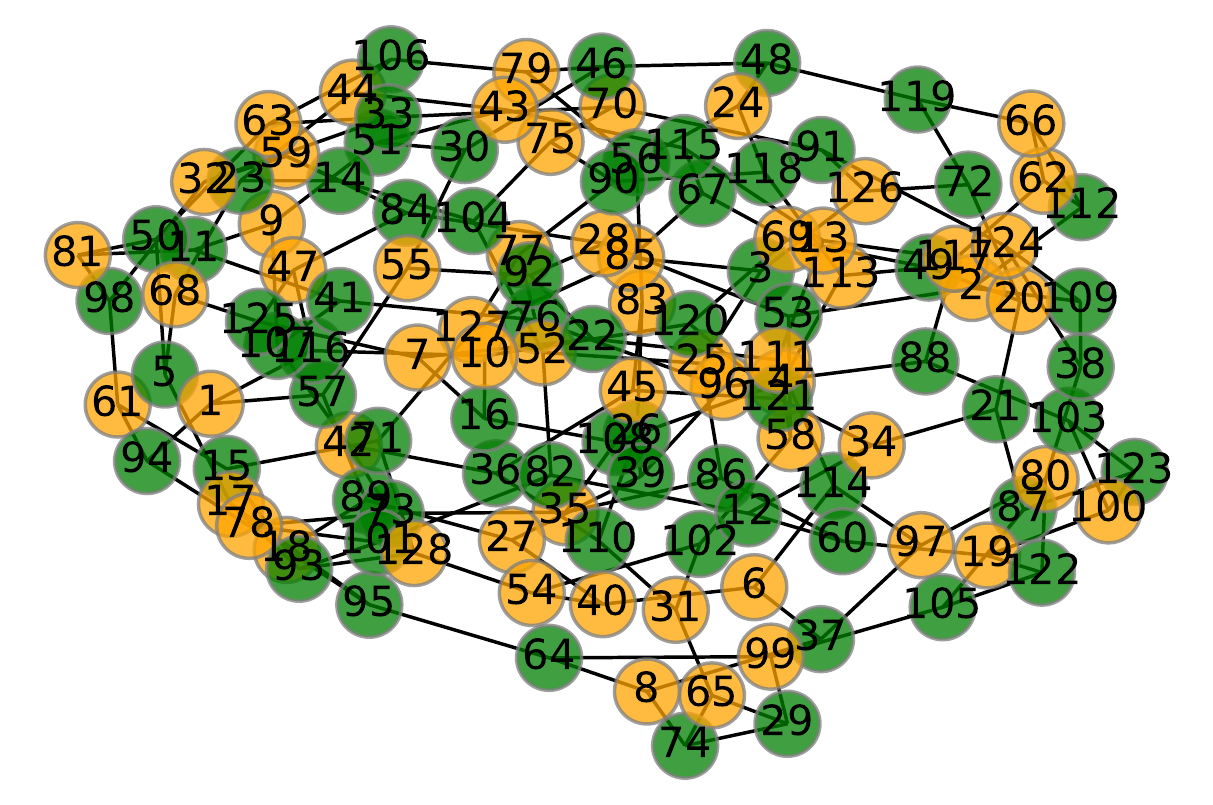}
			\put(0,70){(d) Qiskit \textit{MaxCut} $=137$}
		\end{overpic}
	\end{minipage}
	
	\vspace{1cm}
	
	\begin{minipage}{0.5\textwidth}
		\begin{overpic}[width=1.0\textwidth]{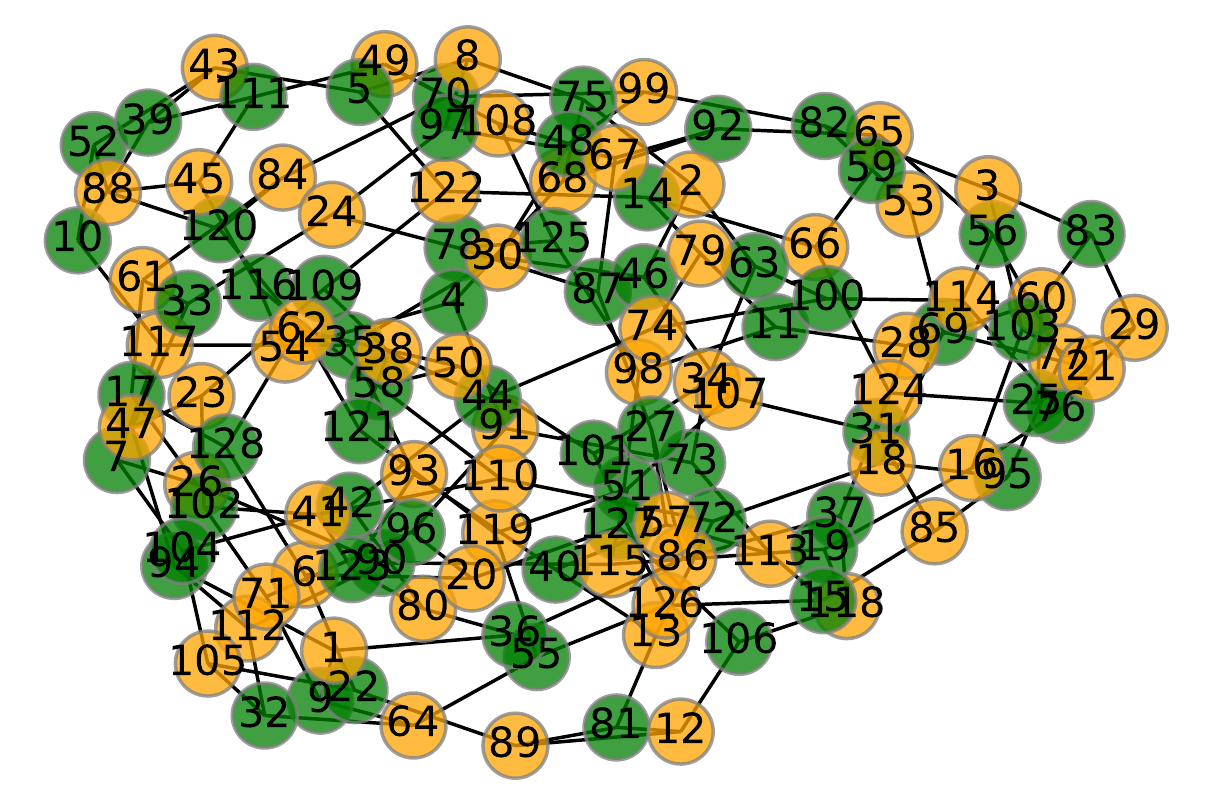}
			\put(0,70){(f) GW \textit{MaxCut} $=166$}
		\end{overpic}
	\end{minipage}\begin{minipage}{0.5\textwidth}
		\begin{overpic}[width=1.0\textwidth]{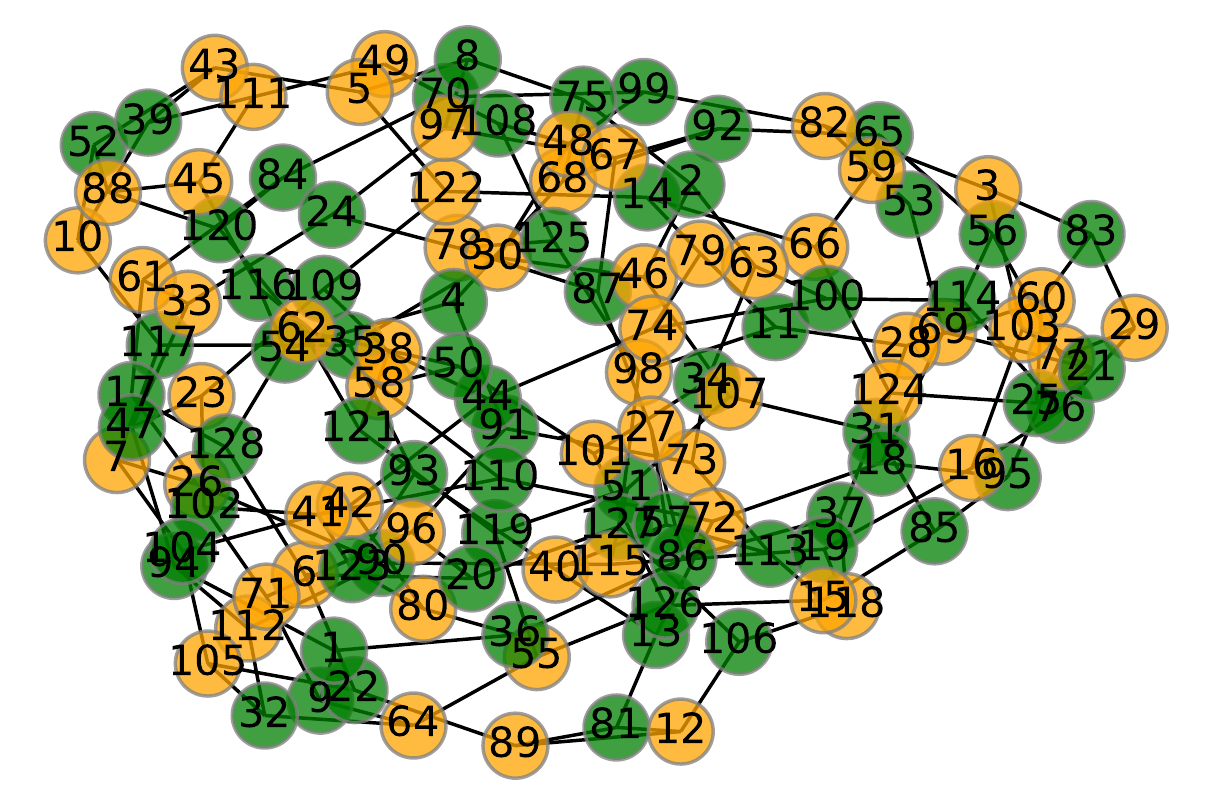}
			\put(0,70){(g)  Qiskit, \textit{MaxCut} $=134$}
		\end{overpic}
	\end{minipage}
	\caption{Randomly generated 3-regular graphs of 128 nodes. Nodes belonging to different partitions are marked in green and orange respectively and the \textit{MaxCut} value is written on top of the graph. Graphs are randomly initialized by Python's networkx package where seed 7 is being used for graphs (a-b), seed 8 for graphs (c-d) and seed 9 for graphs (e-f). GW stands for Goemans-Williamson. \label{fig:cut2}}
\end{figure}

\end{widetext}

\end{document}